\documentclass[pre,twocolumn,superscriptaddress,floatfix,showpacs]{revtex4}

\usepackage[latin2]{inputenc}
\usepackage{amsmath}
\usepackage{amssymb}
\usepackage{epsfig}
\usepackage{color}
\usepackage{bm}
\usepackage{cancel}
\newcommand{\Fn} {F\!\!_{_n}}

\newcommand{\fn} {f\!_{_n}}

\newcommand{\Fnavg}  {\langle F\!\!_{_n} \!\rangle}

\newcommand{\fnaavg} {\bar f\!_{_n}\!(\alpha)}

\newcommand{\Pft}   {\text{P}(f\!_{_t}\!)}
\newcommand{\Pfn}   {\text{P}(f\!_{_n}\!)}
\newcommand{\Pa}    {\text{P}(\alpha)}
\newcommand{\Pfna}  {\text{P}(f\!_{_n},\alpha)}    
\newcommand{\Pfnac} {\text{P}(f\!_{_n}|\alpha)}    

\begin{document}

\title{Anisotropy of force distributions in sheared soft particle systems}

\author{Jens Boberski}
\email{jens.boberski@uni-due.de}
\affiliation{Faculty of Physics, University of Duisburg-Essen, 
D-47048 Duisburg, Germany}

\author{M.\ Reza Shaebani}
\email{shaebani@lusi.uni-sb.de}
\affiliation{Department of Theoretical Physics, Saarland 
University, D-66041 Saarbr\"ucken, Germany}

\author{Dietrich E. Wolf}
\affiliation{Faculty of Physics, University of Duisburg-Essen, 
D-47048 Duisburg, Germany}

\date{\today}

\begin{abstract}
In this numerical study, measurements of the contact 
forces inside a periodic two-dimensional sheared 
system of soft frictional particles are reported. 
The distribution $\Pfn$ of normalized normal forces 
$\fn\!{=}\Fn{/}\Fnavg$ exhibits a gradual broadening 
with increasing the pure shear deformation $\gamma$, 
leading to a slower decay for large forces. The process 
however slows down and $\Pfn$ approaches an invariant 
shape at high $\gamma$. By introducing the joint 
probability distribution $\Pfna$ in sheared configurations, 
it is shown that for a fixed direction $\alpha$, the 
force distribution decays faster than exponentially 
even in a sheared system. The overall broadening can 
be attributed to the averaging over different directions 
in the presence of shear-induced stress anisotropy. 
The distribution of normalized tangential forces almost 
preserves its shape for arbitrary applied strain.
\end{abstract}


\maketitle

{\it Introduction ---}
The contact forces in disordered materials, such as
colloidal suspensions, foams, emulsions, and granular
media are remarkably organized into highly heterogeneous
force networks \cite{Jaeger96}. A statistical mechanical
description of stress transmission in disordered media
should provide a way to understand and predict the
contact force distributions. The tail behaviour of the
normalized normal force distribution $P(\fn\!{\equiv}
\Fn{/}\Fnavg)$ has received much attention, and several 
theoretical models with different assumptions and approaches 
\cite{TheoreticalRefs,Tighe08} predict an exponential 
as well as a Gaussian tail. While early experiments 
and numerical simulations \cite{Radjai96,Silbert02,
Erikson02} favoured the exponential decay, further
studies revealed that the decay can also be faster than
exponential \cite{Makse00,Ohern01,Zhang05,Zhou06,Majmudar05}.
A recent numerical study \cite{Boberski13} of frictional
soft particle systems under pure compression showed that, 
independent of the distance from jamming, the tail behaviour 
can be described by a stretched exponential with an exponent 
around $1.8$, which slightly depends on the choice of the 
contact force law, the friction coefficient, and the 
relative particle stiffness in tangential and normal 
directions.

In sheared systems, a slower decay of $\Pfn$ compared to 
isotropic packings has been observed \cite{Ohern01,Majmudar05,
Snoeijer04,vanEerd07}, where increasing the shear stress 
enhances the broadening of $\Pfn$. This necessitates further 
efforts to provide a comprehensive description of the 
mechanisms underlying stress propagation in sheared systems.
In this Letter, the force distributions in periodic 2D 
granular systems under non-cyclic pure shear are studied. 
The shear-induced stress anisotropy is taken into account 
by categorizing the contacts in terms of their orientation. 
While the normal force distribution decays even faster 
than exponential for the contacts oriented along the 
same direction, it is shown that averaging over all 
angle-resolved distributions leads to the broader shape 
of the overall distribution $\Pfn$ in the sheared system. 
Thus, a connection between the shear-induced stress 
anisotropy and the broadening of $\Pfn$, is established 
which results in the saturation of broadening of $\Pfn$ 
at high shear deformations. In the asymptotic strain-independent 
regime, the distribution for any given direction nearly 
follows a Gaussian form. This enables one to integrate 
over all directions and obtain an approximate analytical 
expression for the invariant broad shape of $\Pfn$ at 
the limit of large shear deformations. 

The distribution $\Pft$ of tangential forces decreases
monotonically (in contrast to $\Pfn$ that usually develops
a peak) with a broad exponential-like tail \cite{Majmudar05,
Silbert02,Hidalgo09,Radjai96,Zhang10,Boberski13}. Moreover,
the collapse of $\Pft$ curves have been reported for different
values of inter-particle friction coefficients \cite{Silbert02},
and for different isotropic \cite{Zhang10,Boberski13} or
anisotropic \cite{Zhang10} applied loads. Here, it is 
verified that the angle-resolved tangential force distributions
nearly collapse onto a universal curve for different orientations.
Therefore, the shape of $\Pft$ remains approximately invariant
with the applied load, in contrast to the distinctive shape 
of the normal force distribution $\Pfn$ under isotropic or 
shear strain. 

\begin{figure}
\centering
\includegraphics[width=0.45\textwidth]{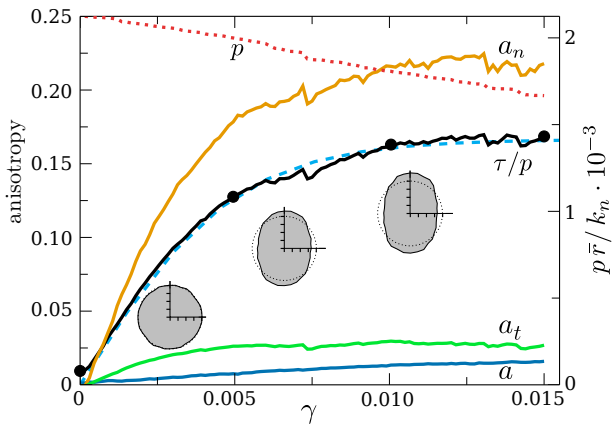}
\caption{The evolution of the fabric and mechanical anisotropies 
($a$, $a_n$, $a_t$) and the stress ratio $\tau{/}p$ (solid lines), 
and the pressure $p$ (dotted line) with increasing the shear strain 
$\gamma$. The dashed line is a fit to the function $\tau{/}p{=}m 
\tanh (\gamma {/}\gamma\!_{_0})$, with $m{\simeq}0.167$ and 
$\gamma\!_{_0}{\simeq}0.005$. The solid circles indicate the 
states for which the force distributions are compared in 
Figs.~\ref{Fig:Normal_M} and \ref{Fig:TangentialF}. Insets: 
The angular distribution of mean normal forces $\fnaavg$ 
before ($\gamma{=}0$) and during ($\gamma{=}0.005$) the 
shear deformation, and when $\tau/p$ saturates 
($\gamma{=}0.01$).}
\label{Fig:stress}
\end{figure}

{\it Numerical Method ---}
The evolution of the contact forces during a quasi-static,
pure shear deformation of an initially compressed packing
of disks was studied numerically. The simulations were
carried out by means of discrete element methods. The
inter-particle forces are modelled by damped, linear 
springs, for both normal and tangential interactions, 
using the spring constant ratio $k_t/k_n{=}0.5$. 
Additionally, the tangential forces obey Coulomb's 
law with a friction coefficient set to $\mu{=}0.5$. The 
two dimensional simulation cell with fully periodic 
boundary conditions contains nearly $20000$ disks. The 
radii are uniformly distributed in the range $[0.8 
\bar r,1.2 \bar r]$. The average particle radius $\bar r$ 
is the length unit, and $k_n \bar r$ is taken as the unit 
of force in the following.

The initial configuration is generated by placing the
particles randomly into the simulation box, without
accepting any overlap between them. This unjammed system
is subsequently compressed quasi-statically by applying
consecutive steps of incremental compression and
relaxation. The compression is achieved by re-scaling
the particle positions while keeping their radii fixed.
The relaxation procedure ensures that the net force
exerted on each particle is 8 orders of magnitude below
the mean contact force. After the average normal overlap
in the jammed state reaches a desired threshold, the
system is sheared in an analogous manner with an applied
pure shear deformation, so that the aspect ratio of the
rectangular simulation box is changed, but the volume is
kept constant. Each time the system is equilibrated, the
force state of the packing is stored. Upon increasing 
shear strain, the fabric and the force network change 
and anisotropies develop. The texture $\Pa$ and the 
average normal and tangential forces can be well 
approximated by a second-order Fourier expansion as 
\cite{Rothenberg89Radjai98} 
\begin{eqnarray}
\begin{aligned}
&\Pa \,\, = \frac{1}{2\pi} \big(1+a \cos(2\alpha)\big),\\
&\bar f_n(\alpha) = 1+a_n \cos(2\alpha),\\
&\bar f_t(\alpha) \, = a_t \sin(2\alpha),
\label{Eq:FourierFits}
\end{aligned}
\end{eqnarray}
where $a$ is the fabric anisotropy, and $a_n$ and $a_t$ 
represent the mechanical anisotropies in normal and 
tangential directions. Figure \ref{Fig:stress} shows 
how the anisotropies develop with increasing the shear 
strain. The relation between the stress ratio and the 
anisotropies follows $\tau{/}p{=}(a{+}a_n{+}a_t)/2$ 
\cite{Rothenberg89Radjai98} with very small deviations 
throughout the shearing process (not shown).

\begin{figure}
\centering
\includegraphics[width=0.45\textwidth]{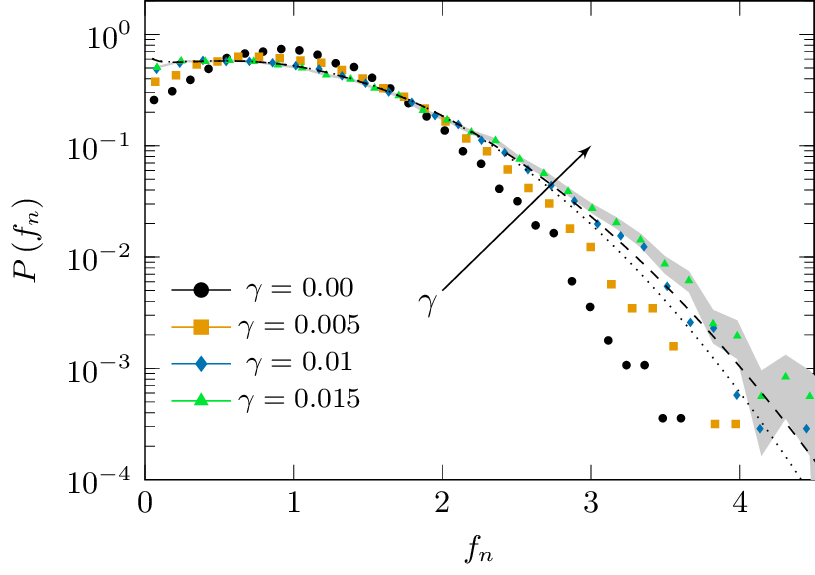}
\caption{The distribution of the normalized normal forces
$\fn{=}\Fn{/}\Fnavg$ for increasing shear strain $\gamma$.
The shaded areas indicate the standard deviation of a 
multinomial distribution to indicate the uncertainty of 
the measured values for $\gamma=0.015$. The dashed line 
indicates the integration over the fitted angle-resolved 
distributions using Eqs.~(\ref{Eq:JointFit}) and 
(\ref{Eq:PfnAsymptotic}) at high $\gamma$ regime. The 
dotted line corresponds to the approximation given by 
Eq.~(\ref{Eq:PfnAsymptoticApprox}). }
\label{Fig:Normal_M}
\end{figure}

The components of the globally averaged stress tensor are
measured using
\begin{equation}
\sigma_{ij} = \frac 1 A \sum_{c=1}^{N_c} f_i^c r_j^c\,,
\label{Eq:stress}
\end{equation}
where $A$ is the area of the system, $f_i^c$ the $i$-th
component of the force acting on contact $c$ and $r_j^c$
the $j$-th component of the branch vector. The sum runs
over all contacts $N_c$ in the system. Denoting the
eigenvalues of the stress tensor by $\sigma_1$ and
$\sigma_2$ ($\sigma_1 {\leq} \sigma_2$), the pressure and
shear stress are given by $p{=}\frac 12 (\sigma_1 {+}
\sigma_2)$ and $\tau {=} \frac 12 (\sigma_2 {-} \sigma_1)$,
respectively. When the isotropic system with aspect ratio
$a{=}1$ is subject to a pure shear deformation, the 
engineering shear strain $\gamma$ increases with 
decreasing the aspect ratio as $\gamma {=} \frac {1-a^2}{2a}$. 
The principal axes of the stress rotate less than $1.13^o$ 
with respect to the biaxial deformation directions 
throughout the shearing process. Figure~\ref{Fig:stress} 
shows that the stress ratio $\tau{/}p$ grows as $\gamma$ 
increases, and eventually saturates for large shear 
strains $(0.01{\leq}\gamma)$. Note, that the shear 
strain at which $\tau{/}p$ and the shear stress 
saturate depend on the volumetric strain applied 
on the initial isotropic packing \cite{Shaebani12}.

\begin{figure}
\centering
\includegraphics[width=0.45\textwidth]{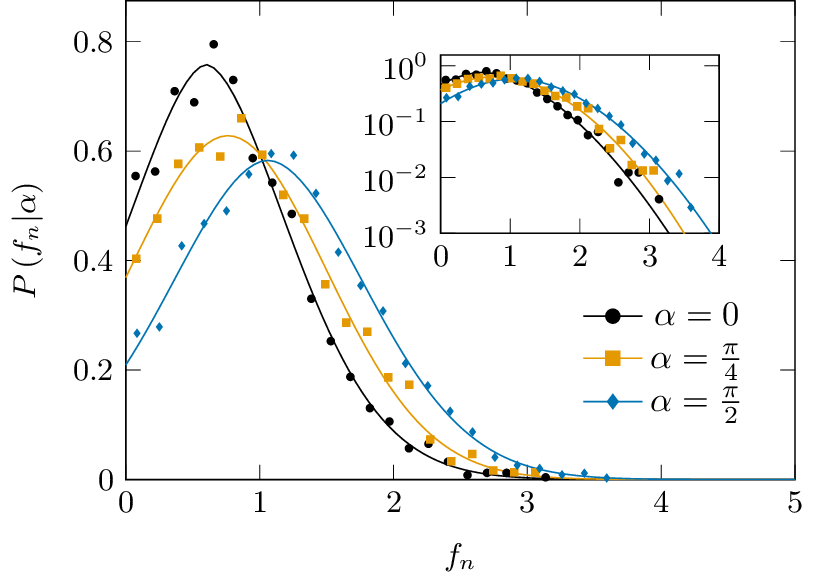}
\caption{The angle-resolved distribution $\Pfnac$ along
three different directions in the packing with $\gamma{=}
0.005$ (symbols). The lines indicate fits given by 
Eq.~(\ref{Eq:fn}).
Inset: The same plot in log-linear scale.}
\label{Fig:Normal_angle_dist}
\end{figure}

{\it Results ---}
Upon increasing the shear deformation $\gamma$, 
the normalized normal force distribution $\Pfn$
broadens, as shown in Fig.~\ref{Fig:Normal_M}. Similar
results were observed in numerical studies
\cite{Makse00,Ohern01,Zhou06} as well as in experiments
with photo-elastic particles \cite{Majmudar05}. A
crucial question is, how the shape of $\Pfn$
is influenced by the characteristics of the globally
imposed stress, namely $p$ and $\tau$. In spite of the
conserved volume during the pure shear deformation,
the pressure can change and one may partially attribute the
shape change of $\Pfn$ at different values of $\gamma$
to the difference between their pressures. Moreover, it
is known that shearing induces anisotropies, leading to
spatial correlations between contact forces with
direction-dependent correlation lengths \cite{Majmudar05}.
While both $p$ and $\tau$ seems to influence the shape
of $\Pfn$, the evolution of $\Pfn$ in
Fig.~\ref{Fig:Normal_M} remarkably slows down at high
$\gamma$ and eventually saturates, which is reminiscent
of the behaviour of stress ratio $\tau{/}p$ and 
anisotropy development in Fig.~\ref{Fig:stress}. Note that, 
when the stress anisotropy approaches an invariant state, 
the shape of $\Pfn$ does not vary any more.

To elucidate the influence of shear-induced stress anisotropy, 
categorizing the contacts according to their orientation 
provides useful information about the angular dependence 
of force transmission. Therefore, the joint probability 
distribution $\Pfna$ for the normal force and the contact 
angle is introduced. The contacts are divided into $12$ 
angular bins of $15^o$ each. Next, the angle-dependent 
conditional distribution $\Pfnac{=}\frac{\Pfna}{\Pa}$
is calculated in each bin, where $\text{P}(\alpha){=}
\int_0^\infty \! \text{P}(f\!_{_n},\alpha) \, \text{d}f\!_{_n}$. 
Three examples 
of the resulting distributions are shown for different 
directions in Fig.~\ref{Fig:Normal_angle_dist}. 
Similar distributions have been recently reported 
in 3D packings under periodic uniaxial shear \cite{Imole14} 
as well as plane shear in a split-bottom Couette cell 
\cite{Singh14}. One finds, that fits of the form
\begin{equation}
\displaystyle\Pfnac=\frac{1}{N(\alpha)} \fn^{\;\nu_n(\alpha)}
\exp\left[-\left| \frac{\fn-b_n(\alpha)}{2 w_n(\alpha)}
\right|^{\delta_n (\alpha)}\right]\,,
\label{Eq:fn}
\end{equation}
characterize the overall shape of $\Pfnac$ along different
directions and for different values of shear strain
$\gamma$. Note that there are only three independent fit
parameters in the above equation, due to the normalization
constraint $\int_0^\infty \!\! \int_0^{2\pi} \Pfna \; 
\text{d}\alpha \, \text{d}\fn{=}1$, and the constraint 
on the first moment of distribution $\int_0^\infty \!\! 
\int_0^{2\pi} \fn \, \Pfna \, \text{d}\alpha \, \text{d}
\fn{=}1$. The choice of force distribution in Eq.~(\ref{Eq:fn}) 
is inspired by the recent work by Tighe et al.\ \cite{Tighe08}, 
where a similar function (even though with slight differences) 
was proposed based on entropy maximization arguments with 
respect to the allowed force network ensemble \cite{Ensemble} 
in isotropic systems. Note that other variants of the fit 
function have been also proposed, see e.g.\ \cite{Muller10} 
for the force distributions in 3D isotropic packings.

\begin{figure}
\centering
\includegraphics[width=0.45\textwidth]{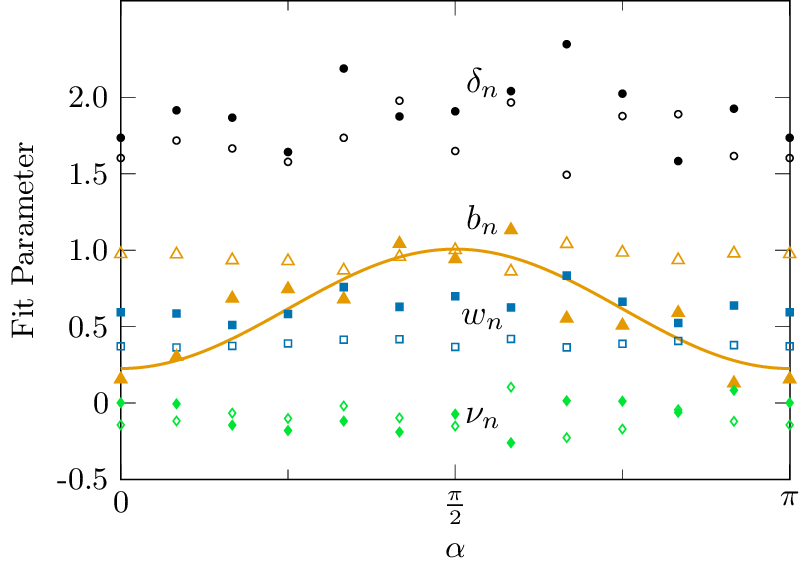}
\caption{The angular dependence of the fit parameters for 
the initial isotropic packing with $\gamma{=}0$ (open symbols) 
and a sheared packing with $\gamma{=}0.015$ (full symbols). 
The fitted line is given by Eq.~(\ref{Eq:Fourier}).}
\label{Fig:Normal_fit_parameters}
\end{figure}

The angular dependence of the fit parameters in the 
initial isotropic packing ($\gamma{=}0$) is compared 
with a highly sheared case ($\gamma{=}0.015$) in 
Fig.~\ref{Fig:Normal_fit_parameters}. One observes 
that the fit parameters are practically $\alpha$-independent 
except for $b_n$ which develops a pronounced angular dependence 
during shearing. This behaviour can be understood by comparing
Eq.~(\ref{Eq:fn}) in the special case of $\delta
{=}2$ with the derivation in ~\cite{Tighe08}. There $w$ and $\nu$ are related to the local force 
balance constraint on the grains, the friction coefficient, 
and the connectivity of the force network, thus, they are 
not expected to be angle dependent. On the other hand, 
$b_n$ is set by a constraint to the pressure. Since the average normal force varies with 
$\alpha$ in the presence of stress anisotropy in the system, 
it becomes clear why $b_n$ is $\alpha$-dependent. However, 
note that pressure is not the only control parameter 
in determining the shape of the angle-resolved distributions. 
$\text{P}(f\!_{_n}|\alpha)$ along a given direction in 
the sheared system notably differs from $P(f_n)$ of an 
isotropic packing carrying the same average normal force ~\cite{Boberski13}. 
Our generalized 
Eq.~(\ref{Eq:fn}) captures the shape of sheared force 
distributions by allowing the exponent $\delta$ to act 
as an additional free parameter. Nevertheless, the global 
constraint on the applied shear stress/strain has to 
be taken into account to obtain an analytical expression 
for $P(f_n)$ in sheared packings.

\begin{figure}
\centering
\includegraphics[width=0.45\textwidth]{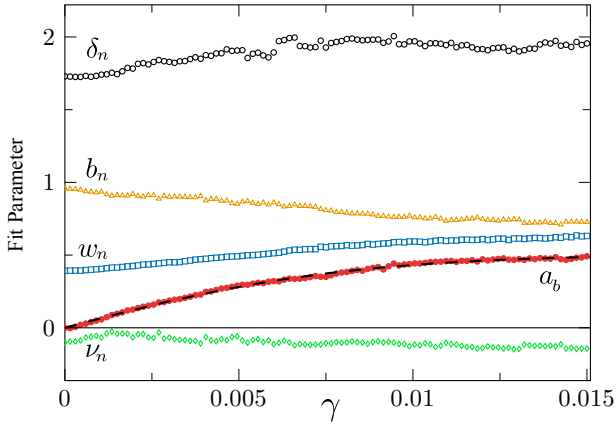}
\caption{The evolution of the fit parameters and the anisotropy 
$a_{_b}$ with shear strain $\gamma$. The dashed line is a 
fit to the function $a_{_b}{=}m\tanh (n\gamma)$, with 
$m {\simeq} 0.51$ and $n {\simeq} 124.5$.}
\label{Fig:Normal_aniso}
\end{figure}

The parameter $b_n$ varies periodically with 
a peak in the direction of compression, which can be 
described by a second-order Fourier expansion of the form
\begin{equation}
b_n(\alpha) = b_n \big[ 1 {-} a_{_b} \cos(2\alpha)\big].
\label{Eq:Fourier}
\end{equation}
This equation and the finding that the rest of fit 
parameters do not develop a clear angular dependence during 
shearing motivates us to propose a similar 
functional form for the joint probability distribution $\Pfna$, 
assuming that only the shift parameter $b_n$ has an angular 
dependence according to Eq.~(\ref{Eq:Fourier}), i.e.\ 
\begin{equation}
\Pfna=\frac{1}{N} \fn^{\;
\nu_n} \exp\!\!\left[-\left| \frac{\fn- b_n \big[ 1 
{-} a_{_b} \cos(2\alpha) \big]}{2 w_n} \right|^{\delta_n}
\right].
\label{Eq:JointFit}
\end{equation}
The results of the fits 
via Eq.~(\ref{Eq:JointFit}) are shown in
figure \ref{Fig:Normal_aniso} by the evolution of the
fit parameters of Eq.~(\ref{Eq:JointFit}) for increasing 
$\gamma$. Note, that they approach an invariant state 
for large shear strains.

\begin{figure}
\centering
\includegraphics[width=0.45\textwidth]{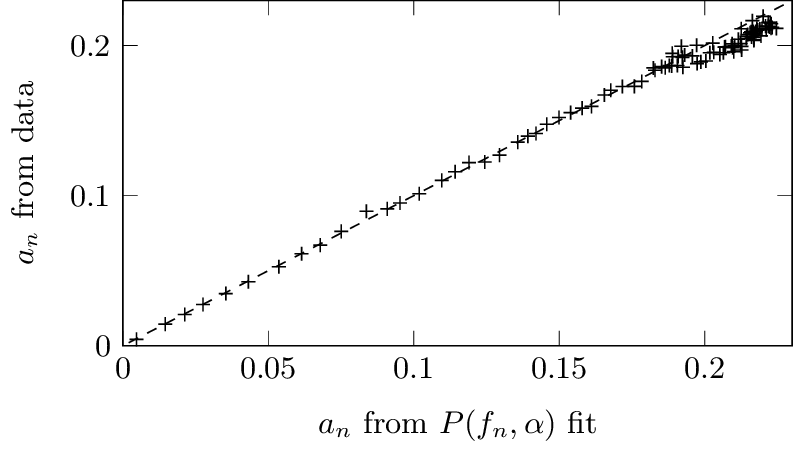}
\caption{The mean normal force anisotropy $a_n$ of the 
simulation data vs.\ the one obtained from integrating 
the joint distribution via Eq.~(\ref{Eq:PfnAsymptotic}). 
The dashed line indicates identity.}
\label{Fig:FitFit}
\end{figure}

By integrating the regulated form of $\Pfna$ in 
Eq.~(\ref{Eq:JointFit}) over $\alpha$, one obtains the 
overall distribution  
\begin{eqnarray}
\begin{aligned}
&\Pfn = \int_0^{2\pi} \!\!\! \Pfna \; \mathrm d \alpha, 
\end{aligned}
\label{Eq:PfnAsymptotic}
\end{eqnarray}
which can be compared to $\Pfn$ obtained from the simulations. 
Using numerical integration (due to the non-integer exponent 
$\delta$), $\Pfn$ was obtained, e.g. for the packing with 
$\gamma{=}0.015$. The resulting curve, shown in 
Fig.~\ref{Fig:Normal_M}, matches the simulation results.
It was also checked, whether the resulting $\Pfn$ 
reproduces the anisotropy $a_n$ obtained directly from 
the simulation data. Figure~\ref{Fig:FitFit} 
shows that both anisotropies are in good agreement, 
with small deviations for large anisotropies. 

Interestingly, the exponent of the stretched 
exponential $\delta_n$ increases during shearing and 
approaches two, \textit{i.e.} it nearly follows a Gaussian 
tail at high $\gamma$ [see Fig.~\ref{Fig:Normal_aniso}]. 
This allows one to analytically integrate the angle-resolved 
distribution (i.e.\ by combining 
Eqs.~(\ref{Eq:JointFit}) and (\ref{Eq:PfnAsymptotic}) 
using $\delta{=}2$) and obtain an approximate 
$\gamma$-invariant expression for the marginal 
distribution $\Pfn$ in the limit of large shear strains
\begin{eqnarray}
\lim_{\gamma \to \infty}\!\!\Pfn {\approx} \frac{2\pi}{N} 
\fn^{\;\nu} \exp\!\left[-\Big(\frac{\fn \! {-} 
b\!_{_n}}{2w}\Big)^{\!2}\right] \text{I}_0 \Big(\frac{a_{_b} 
b\!_{_n} (b\!_{_n}{-}\fn)}{2w^{\,2}}\Big),
\label{Eq:PfnAsymptoticApprox}
\end{eqnarray}
where $\text{I}_0$ is the modified Bessel function of 
the first kind. In the integration, all quadratic terms 
in $a _{_b}\cos(\alpha)$ are neglected. The above expression is 
compared to the simulation data in Fig.~\ref{Fig:Normal_M}, 
which shows a satisfactory agreement. As expected, the 
decay is slightly faster than simulations, since a pure 
Gaussian exponent is used to obtain Eq.~(\ref{Eq:PfnAsymptoticApprox}) 
despite the fact that $\delta_n$ converges to an exponent 
slightly below $2$. 

Note that after extensive slow shearing, a system is expected 
to reach a critical state of flow, where the force state 
attains a statistically steady-state condition. However, we
limit the application range of our results to quasi-static 
deformations because yielded systems may behave differently, 
as the flow properties are in general shear-rate dependent 
\cite{Corwin05}. Investigation of granular flows is 
beyond the scope of this letter.

\begin{figure}[b]
\centering
\includegraphics[width=0.45\textwidth]{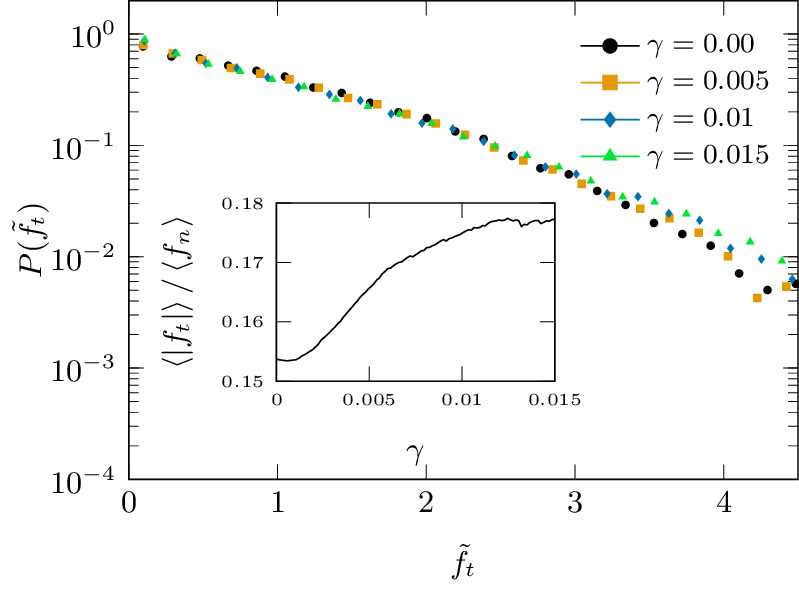}
\caption{The distribution of normalized tangential forces 
$P(\tilde f_t)$ for different values of shear strain, 
$\gamma$. Inset: The evolution of $\langle |f_t| \rangle 
/ \langle f_n \rangle$ with increasing $\gamma$.}
\label{Fig:TangentialF}
\end{figure}

The marginal distribution of the tangential forces $P(\tilde f_t)$, 
using $\tilde f_t = |f_t|/\left<|f_t|\right>$, decreases 
monotonically, as shown in Fig.~\ref{Fig:TangentialF}. There 
is no significant change of the distribution during shearing. 
This result, together with the fact that the anisotropy 
$a_t$ of the average tangential forces is about an 
order of magnitude below the anisotropy $a_n$ of 
the average normal forces in the sheared system (see 
Fig.~\ref{Fig:stress}), shows that while friction stabilizes 
a packing at a lower coordination number, the main history 
dependence of the contact forces is observed in the 
normal forces.

{\it Conclusion ---}
The normalized contact force distributions in sheared systems of 
soft frictional particles were studied numerically. 
The broad shape 
of the marginal distribution $\Pfn$ can be attributed to 
averaging over different contact orientations, which 
carry different stresses. While the angle-resolved distribution along an 
arbitrary direction decays faster than exponential 
similar to the behaviour of the isotropic packings. 
However, integration over all directions modifies 
the shape of overall distribution $P(f_n)$ in a sheared system at large forces, 
resulting in a broad 
distribution with an apparent slower decay. 
Therefore, a
link between the broadening of the normalized normal 
force distribution $\Pfn$ and the shear-induced 
stress anisotropy was established. 

The broadening is enhanced with 
increasing shear strain, as far as the stress 
anisotropy still develops. Eventually, the stress 
anisotropy saturates at high shear deformations, 
thus, $\Pfn$ reaches a strain-independent shape. 
Since the anisotropy development is considerably 
weaker in the tangential direction, $\Pft$ remains 
approximately invariant through the deformation 
process. The fabric anisotropy 
remains small throughout the shearing process 
and has no major influence on the shape of the 
force distributions.

These findings show that the stress propagation 
in sheared systems can be better understood when angle-resolved 
distributions are considered. Analytical treatments 
\cite{Tighe08} to obtain the force distributions in isotropic 
packings need to be reconsidered in sheared systems by taking 
the global constraint on the applied shear stress/strain into 
account, as an step forward towards fully describing the shape 
of force distributions in sheared packings. The results also 
help to better understand the mechanisms of deformation of 
granular materials at the microscopic level, which 
facilitates the development of stochastic approaches 
\cite{Saitoh14} for theoretical modelling of deformation 
and elastic behaviour of granular systems.

We would like to acknowledge the support by the German
Research Foundation (DFG) via priority program SPP 1486 
``Particles in Contact" and the Center for Computational
Sciences and Simulation of the University of Duisburg-Essen.

\end{document}